\documentclass[9pt,twocolumn,twoside]{opticajnl}
\journal{opticajournal}

\setboolean{shortarticle}{false}
\setboolean{displaycopyright}{false}

\usepackage{amsmath,bm}
\usepackage{booktabs}
\usepackage{subcaption}
\usepackage{siunitx}
\usepackage{xspace}
\usepackage{lineno}
\usepackage{pifont}
\usepackage{stfloats}
\usepackage{placeins}

\graphicspath{{./figures/}}
\allowdisplaybreaks[2]
\title{Delta-sigma noise shaping for low-resolution quantization in Gaussian-modulated continuous-variable quantum key distribution}

\author[1]{Wenjun Fan}
\author[1]{Zhenlin Zhao}
\author[1,*]{Dawei Wang}
\affil[1]{Guangdong Provincial Key Laboratory of Optoelectronic Information Processing Chips and Systems, School of Electronics and Information Technology, Sun Yat-sen University, Guangzhou, 510006, China}
\affil[*]{Corresponding author: wangdw9@mail.sysu.edu.cn}
\doi{}

\begin{abstract}
Low-resolution digital-to-analog conversion can simplify continuous-variable quantum key distribution (CVQKD) transmitters, but coarse quantization increases excess noise and can generate data-correlated spectral replicas.
We demonstrate digital delta--sigma modulation (DSM) for low-resolution waveform generation in Gaussian-modulated CVQKD.
In a 500-MBaud CVQKD system over 20~km of fiber, 3-bit second-order DSM achieves an estimated asymptotic secret key rate of 46.38~Mbps at an oversampling ratio of 12, whereas uniform quantization yields no positive rate.
With 4-bit second-order DSM, the rate reaches 50.11~Mbps, approximately 97\% of the 12-bit reference.
A noise model combining the DSM noise-transfer function with matched filtering guides the sampling-rate and signal-band design for effective noise suppression.
We also identify a phase-conjugate spectral replica arising from third-order intermodulation between the pilot tone and quantum signal, and evaluate its key-rate penalty using a linear-replica security model.
At 3-bit resolution and an oversampling ratio of 24, first-order DSM reduces the experimentally estimated leakage-mode strength by approximately four orders of magnitude.
Together, our findings establish the security boundaries and operating conditions of DSM, providing design guidance for low-resolution CVQKD transmitters.
\end{abstract}

\begin{document}

\maketitle


\begin{figure*}[!t]
\centering
\includegraphics[width=0.9\textwidth]{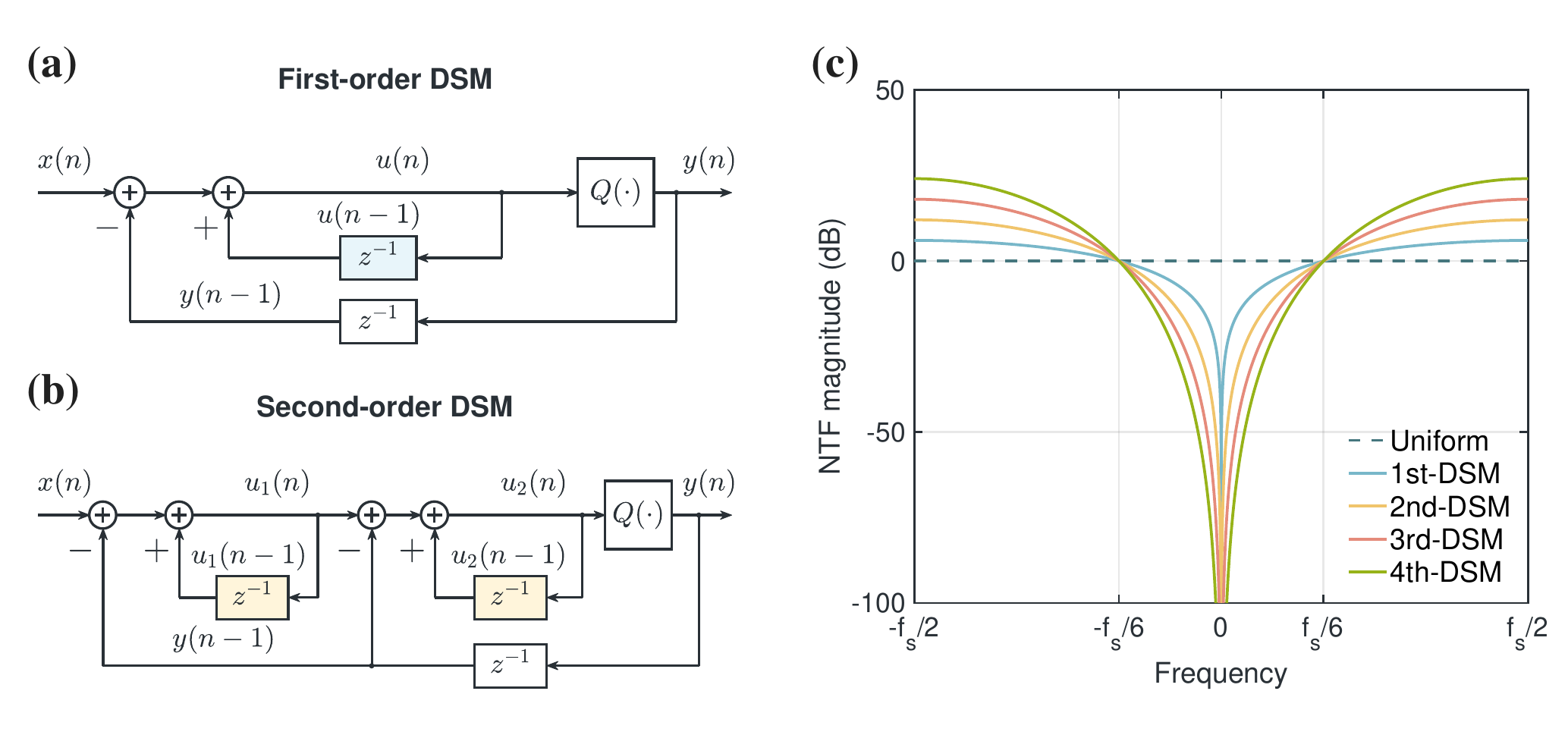}
\caption{
Structure of (a) first-order DSM, (b) second-order DSM, and (c) ideal noise transfer function.
}
\label{fig:dsm}
\end{figure*}

\section{Introduction}

High-speed continuous-variable quantum key distribution (CVQKD) increasingly relies on digital waveform generation and coherent optical communication components, including high-speed digital-to-analog converters (DACs), IQ modulators, balanced photodetectors, and integrated coherent receivers~\cite{Weedbrook2012, Laudenbach2018, Pirandola2020, Zhang2019, Zhang2024Review, Wang2024Telecom, Hajomer2024, Hajomer2026Chip16GBaud}.
Recent silicon photonic implementations further demonstrate the potential for compact transmitters and receivers, including Gaussian-modulated systems with integrated receivers~\cite{Liu2025SiliconTRX, Xu2026GaussianIntegrated}.
At a Gaussian-modulated transmitter, the intended electrical waveform must be represented by the finite amplitude levels available from a practical DAC~\cite{Pietri2024QOSST}.
The difference between the true waveform and the discretized waveform is defined as quantization noise.
At high resolution, quantization error is often approximated as additive white noise independent of the original signal~\cite{Widrow1996} and is usually treated as part of the preparation noise in CVQKD.
Lower DAC resolution reduces transmitter complexity, power consumption, and implementation cost, especially at high speeds~\cite {Almonacil2020}.
However, coarse quantization can distort the prepared states and increase excess noise, making transmitter resolution an important implementation constraint~\cite{Jouguet2012, Kumar2020, Wang2025Composable}.
In pilot-assisted systems, sharing the DAC dynamic range between a weak quantum signal and a stronger phase-reference tone can further increase the required resolution~\cite{Qi2015, Kleis2017, Ren2021, Chin2021CarrierRecovery, Sarmiento2025OpticalPilot}.
Therefore, it is important to understand and mitigate quantization effects while retaining the practical advantages of low-resolution waveform generation.

Existing security analyses show that a sufficiently fine discrete approximation can reproduce ideal Gaussian modulation in practice \cite{Jouguet2012}, while discrete-modulation protocols using a finite constellation offer an alternative route to simpler implementations \cite{Lin2019, Guo2025BasisEncoding, Pan2026DM126km}.
In practical CVQKD implementations, researchers usually incorporate quantization error into the noise budget as an equivalent excess-noise contribution \cite{Kumar2020, Wang2022, Ren2021, Hajomer2024LLO100km}.
Recent simulations comparing electrical and optical pilot generation have also examined low-resolution DAC distortion and shown that optical pilot generation can relax the DAC-resolution requirement \cite{Sarmiento2025OpticalPilot}.
More recently, researchers have jointly optimized transmitter and receiver filters under finite DAC/ADC resolution and bandwidth constraints, retaining quantization error as an additive variance in the system model \cite{Matsenko2026}.
These approaches quantify quantization error but do not explicitly design its statistical features, such as its spectral distribution.

The spectral structure of quantization noise becomes important at medium-to-low resolutions.
Without dither, coarse quantization error need not be white or independent of the modulation data and can produce deterministic, data-correlated spectral components \cite{Gray1993Dither, Widrow1996}.
Coarse quantization is a nonlinear amplitude-mapping operation and can generate harmonic and intermodulation products correlated with the original data \cite{Eielsen2017}.
If these components transfer to unintended optical modes accessible to an eavesdropper, in-band excess-noise variance alone cannot characterize the resulting side channel \cite{Derkach2017, Jain2021}.
The general security implications of unintended modes that carry correlated modulation and are accessible to an eavesdropper have been studied \cite{Derkach2017, Jain2021, Wang2022, Hajomer2022}, but they do not cover the low-DAC-resolution scenario.
A more relevant work shows that sidelobes from the DAC zero-order hold can form a side channel when excluded from calibration and parameter estimation~\cite{Liu2026}.
But the mechanism differs from the low-resolution quantization replicas considered here in their physical origins and spectral locations.
Zero-order-hold images are periodic copies of the sampled spectrum, with locations fixed by the DAC sampling rate.
By contrast, we show that the data-correlated modes from coarse DAC quantization may fall inside the first Nyquist zone.
This distinction motivates examining both in-band noise and data correlations in quantization-induced spectral replicas.

In this work, we investigate transmitter-side quantization noise shaping to suppress both excess noise and data-correlated side channels caused by insufficient DAC resolution.
In particular, we use delta--sigma modulation (DSM) as a simple realization of noise shaping, which uses oversampling and low-order feedback to implement a prescribed noise transfer function (NTF) that suppresses quantization noise in a target band and redistributes it to higher frequencies~\cite{Inose1962, Gray1987, Pavan2017}.
DSM and related error-feedback structures have been used for low-resolution waveform generation, mobile fronthaul, and digital predistortion in classical optical links~\cite{Wang2019DSM, Zeng2023, Xie2024, Yang2025, Zhong2025}.
For CVQKD, the benefit must be evaluated within the effective quantum-signal mode selected by pulse shaping, detection, and digital filtering~\cite{Chen2023ContinuousMode}.

Building on our preliminary results \cite{Fan2026OECC}, this work incorporates the DSM into the quantum transmitter and evaluates quantization-noise shaping in a complete CVQKD experimental system.
We derive an in-band quantization-noise and signal-to-quantization-noise ratio (SQNR) model jointly determined by the NTF, the root-raised-cosine (RRC) pulse-shaping, and matched-filter responses.
We further evaluate the effect of low-resolution quantization in a 500-MBaud Gaussian-modulated CVQKD system over 20~km of fiber.
Experimental results identify the operating region set by quantizer resolution, oversampling ratio (OSR), and DSM order.
At 12-fold oversampling, a 3-bit second-order DSM reduces excess noise from 0.86~SNU (uniform quantization) to 0.028~SNU and restores a positive estimated asymptotic secret key rate of 46.38~Mbps.
With 4-bit second-order DSM, the estimated key rate reaches 50.11~Mbps, approximately 97\% of the 51.66~Mbps reference obtained from the 12-bit waveform generator.
In addition, we find that a first-order DSM reduces the linear correlation between the Gaussian-modulated signal and the quantization-induced replicas by three orders of magnitude relative to the 3-bit uniform quantization.
We also confirm that a proper analog or digital filter can further reduce the amplitude of the quantization-induced replicas.
These results establish a practical operating region for low-resolution Gaussian-modulated CVQKD while demonstrating suppression of quantization-induced linear correlations outside the signal band.

\begin{table*}[!t]
\centering
\caption{Simulation and experimental parameters.}
\label{tab:parameters}
\begin{tabular}{p{0.25\textwidth}p{0.31\textwidth}p{0.31\textwidth}}
\toprule
Parameter & Simulation & Experiment \\
\midrule
Protocol/test waveform & Gaussian modulation; QPSK for spectral-replica diagnostics & Gaussian-modulated LLO-CVQKD \\
Symbol rate & 500~MBaud & 500~MBaud \\
RRC roll-off factor & 0.1 & 0.1 \\
Quantum-signal center frequency & 300~MHz by default; 0.3--2~GHz sweep & 300~MHz \\
Pilot frequency & 700~MHz & 750~MHz \\
Sampling rate/OSR & 1.5--24~GHz / 3--48 & 2--12~GHz / 4--24 \\
Imposed quantizer resolution & 3 and 8~bits; 2, 3, 4, 5~bits used for replica diagnostics & 3--8~bits \\
Fiber and receiver & --- & 20-km SSMF; 1.6-GHz balanced detector \\
Detection parameters & --- & $\eta=0.72$, $v_{\mathrm{el}}=0.076$~SNU \\
Key-rate parameters & --- & $\beta=0.98$, $T=0.355$ \\
\bottomrule
\end{tabular} 
\end{table*}

\begin{figure}[!tbp]
\centering
\includegraphics[width=\linewidth]{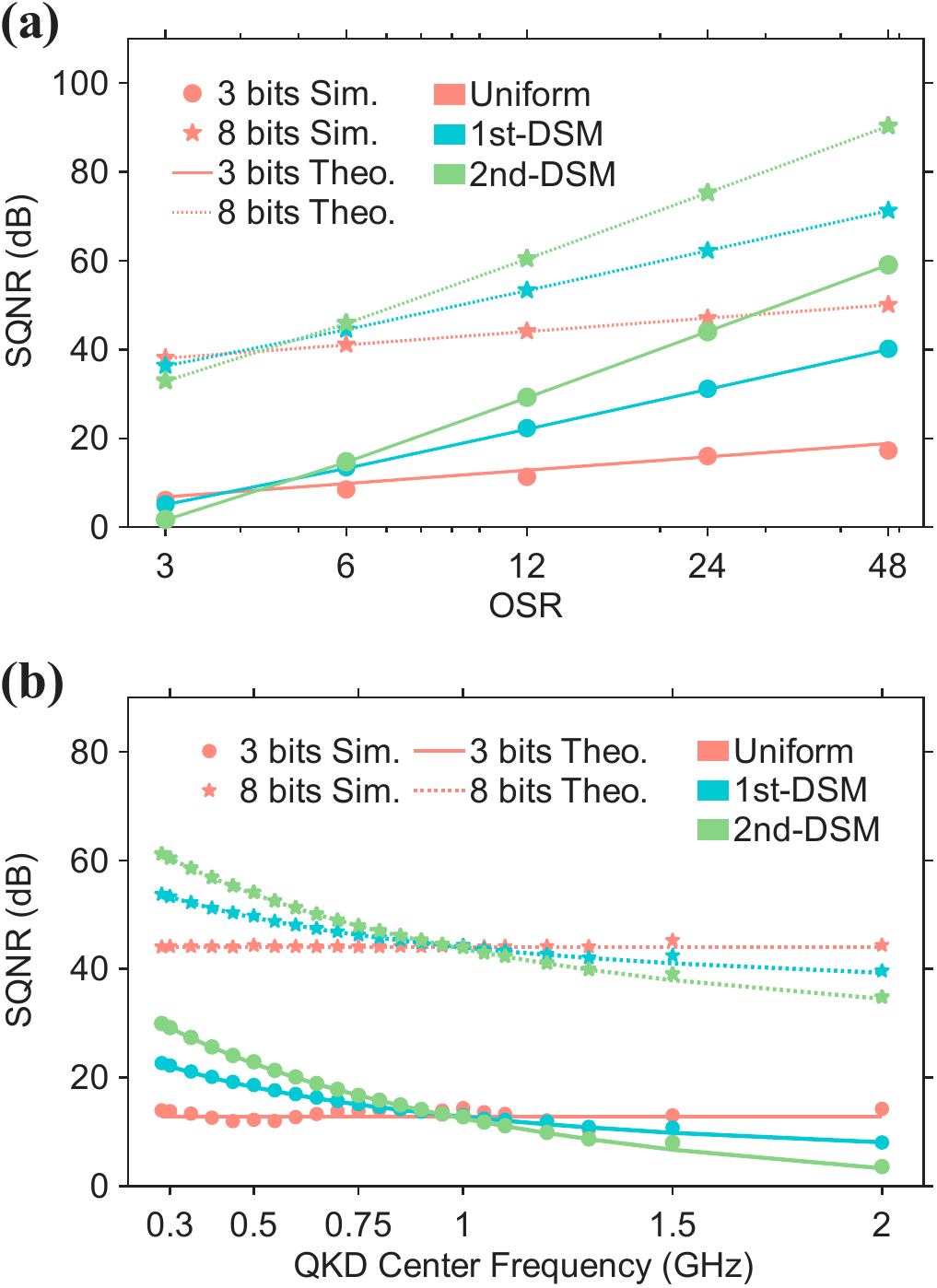}
\caption{
Simulated in-band SQNR and the theoretical prediction based on ~\eqref{eq:sqnr_final}.
(a) SQNR versus OSR ($f_c=300$~MHz);
(b) SQNR versus quantum-signal center frequency (OSR = 12).
}
\label{fig:sqnr}
\end{figure}

\section{DSM Principle and Quantization-Noise Model}

Here, we consider frequency pilot-tone-assisted Gaussian-modulated CVQKD \cite{Pi2023LLO100km}.
Alice generates independent and identically distributed Gaussian variables.
After upsampling and RRC pulse shaping, they form the baseband quantum signal $s_{\mathrm{q}}[n]$ \cite{Pietri2024QOSST}.
Upconverting $s_{\mathrm{q}}[n]$ to the center frequency $f_{\mathrm{c}}$ and combining it with a pilot of frequency $f_{\mathrm{p}}$ and complex amplitude $A_{\mathrm{p}}$ gives the overall waveform
\begin{equation}
\begin{aligned}
x[n]={}&s_{\mathrm{q}}[n]
e^{j2\pi f_{\mathrm{c}}n/F_s}
+A_{\mathrm{p}}
e^{j2\pi f_{\mathrm{p}}n/F_s},
\end{aligned}
\label{eq:composite}
\end{equation}
where $F_s$ is the DAC sampling rate with unit sample-per-second (Sa/s).
Let $R_s$ be the symbol rate, and define the OSR as $F_s/R_s$.

The I and Q components of the composite waveform are quantized independently.
For a bit depth $b$, the quantization steps are
\begin{align}
\Delta_I=A_{I}/(2^b-1), \qquad 
\Delta_Q=A_{Q}/(2^b-1),
\label{eq:quant_step}
\end{align}
where $A_{I}$ and $A_{Q}$ denote the respective full-scale amplitudes.
Let $Q(x)$ denote the uniform quantizer, which maps the waveform $x$ to $2^b$ discrete levels.

Figure~\ref{fig:dsm}(a) and (b) are the block diagrams of first- and second-order DSMs with cascade-of-integrators-with-feedback (CIFB) structures.
The first-order DSM satisfies
\begin{align}
u[n]&=u[n-1]+x[n]-y[n-1],\nonumber\\
y[n]&=Q\!\left(u[n]\right),
\label{eq:dsm1}
\end{align}
and the second-order DSM satisfies
\begin{align}
u_1[n]&=u_1[n-1]+x[n]-y[n-1],\nonumber\\
u_2[n]&=u_2[n-1]+u_1[n]-y[n-1],\nonumber\\
y[n]&=Q\!\left(u_2[n]\right).
\label{eq:dsm2}
\end{align}
The internal quantization error is defined as $e[n] = y[n]-u[n]$ and $e[n] = y[n]-u_2[n]$ for the first- and second-order DSMs, respectively.
Under the linearized additive quantization-error model \cite{Gray1987, Pavan2017}, the signal transfer function (STF) of the $L$th-order DSM is unity, whereas the noise transfer function (NTF) is given by (detailed in the Appendix)
\begin{equation}
\begin{aligned}
\mathrm{NTF}_L(f)
&=
\left[
2j\,e^{-j\pi f/F_s}
\sin\!\left(\frac{\pi f}{F_s}\right)
\right]^L.
\end{aligned}
\label{eq:ntf}
\end{equation}
Let $q[n]=y[n]-x[n]$ denote the quantization error at the DSM output with a power spectral density (PSD) $S_q(f)$.
Let $S_e(f)$ be the PSD of the internal quantization error.
The input--output relation of the DSM gives $S_q(f) = \left|\mathrm{NTF}_L(f)\right|^2 S_e(f).$

Figure~\ref{fig:dsm}(c) plots the quantization-error power gain.
The dashed 0-dB line represents uniform quantization without noise shaping.
For $|f|<F_{s}/6$, the NTF magnitude is below unity and decreases as the DSM order increases, providing progressively stronger suppression near zero frequency.
Note that all orders have a zero at DC and cross the 0-dB reference at $|f|=F_{s}/6$.
Increasing the order therefore strengthens low-frequency suppression without extending the frequency range of quantization noise suppression.
For $F_s/6<|f|\leq F_s/2$, the NTF magnitude exceeds unity, and higher-order DSM produces stronger amplification of the quantization error.
Thus, the benefit of increasing the DSM order depends on the position and signal bandwidth relative to these noise-suppression and amplification regions.
For the frequency-shifted signal considered here, the resulting in-band noise power must be evaluated by weighting the shaped error spectrum with the receiver matched-filter response.

\begin{figure}[!tbp]
\centering
\includegraphics[width=\linewidth]{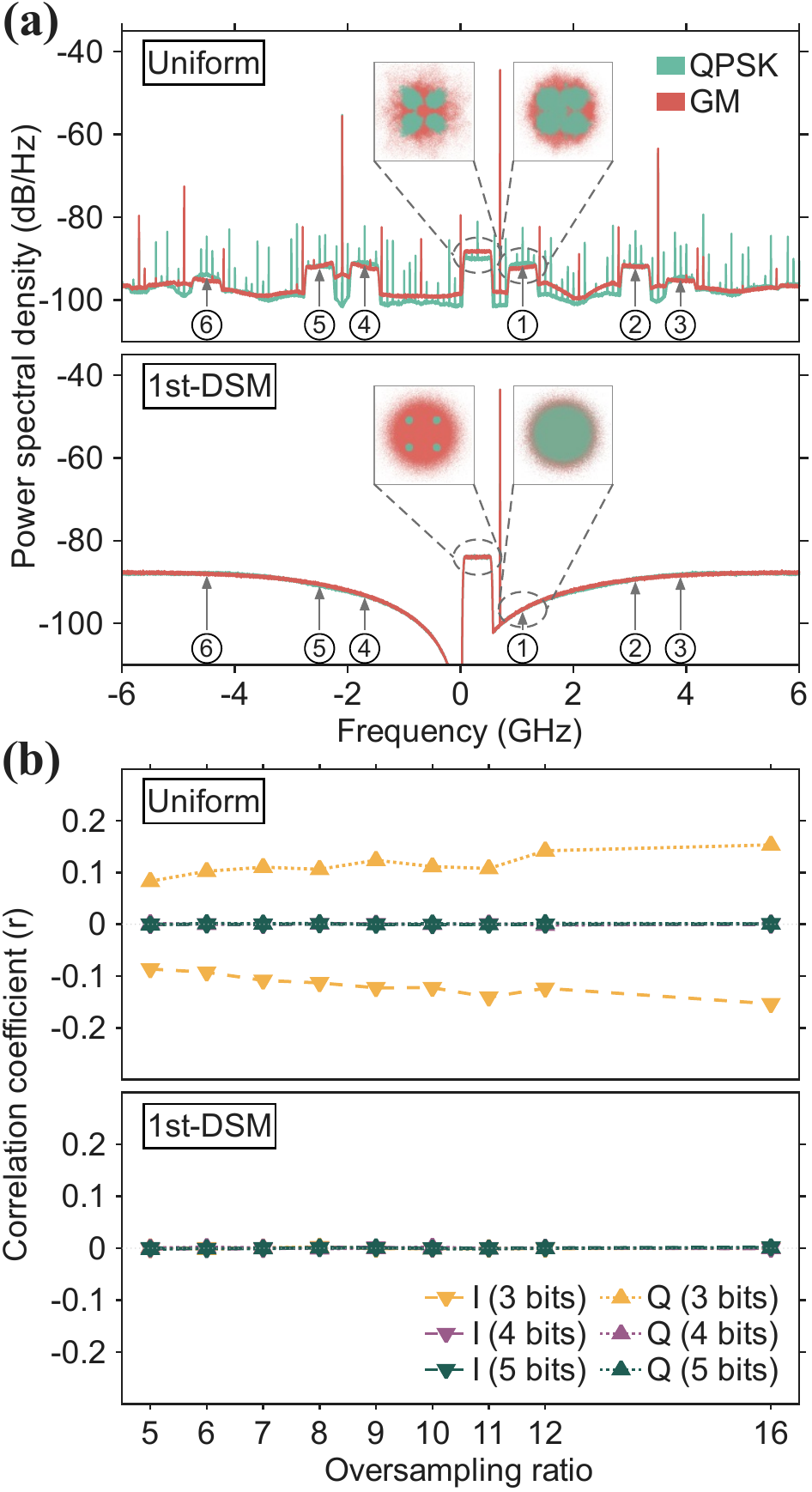}
\caption{
(a) Power spectral densities of QPSK and Gaussian-modulated waveforms under 2-bit quantization. Insets show the constellations of the main signal and the 1.1-GHz replica.
(b) Correlation coefficients between the I and Q components of the 1.1-GHz replica and the corresponding components of the original test symbols versus oversampling ratio.
}
\label{fig:replica}
\end{figure}

\begin{figure}[!tbp]
\centering
\includegraphics[width=\linewidth]{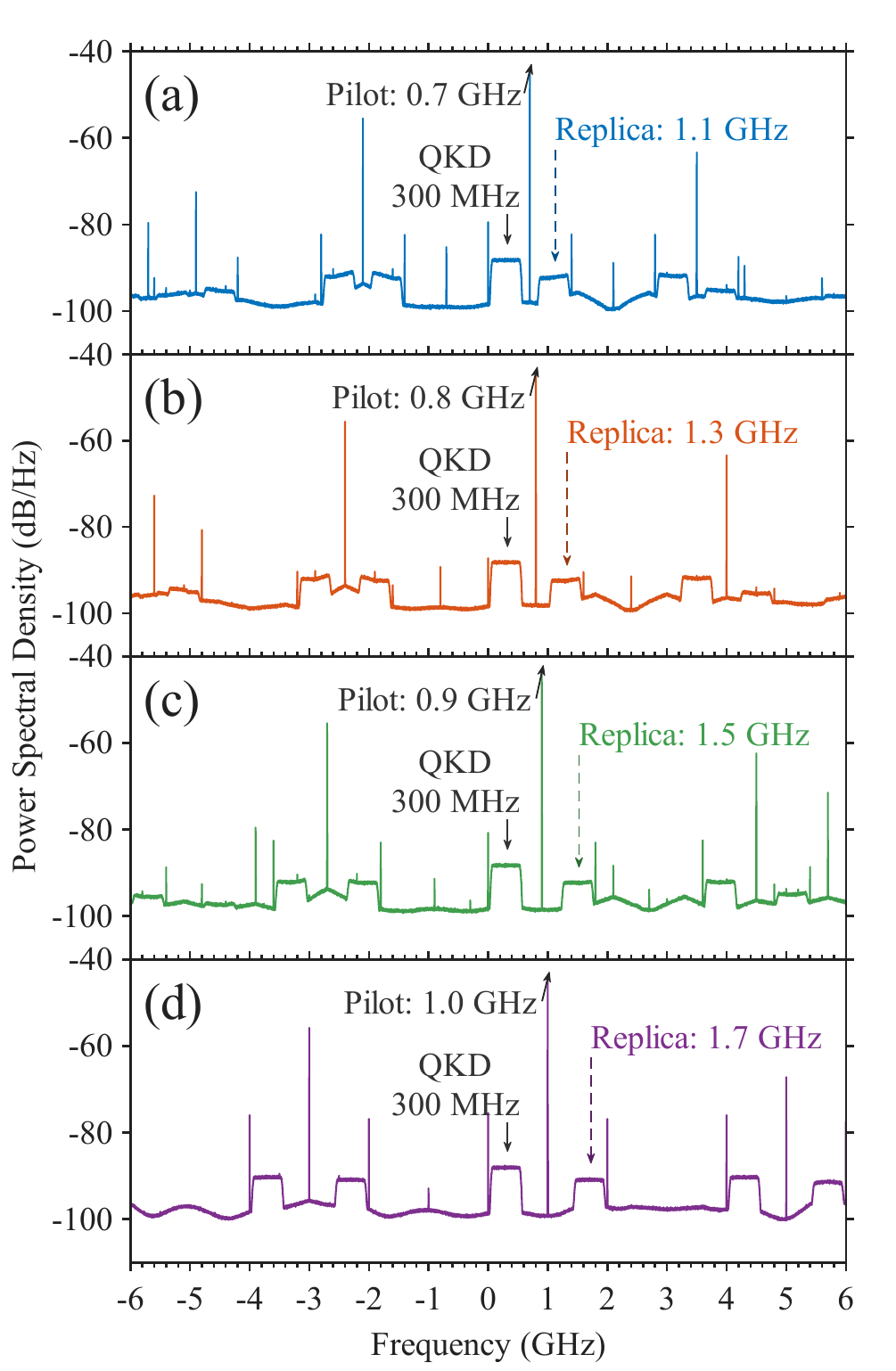}
\caption{
Simulated power spectral densities for different pilot frequencies.
The quantum-signal center frequency is $f_c=300$~MHz, and the sampling rate is 12~GHz.
(a) $f_p=700$~MHz;
(b) $f_p=800$~MHz;
(c) $f_p=900$~MHz;
(d) $f_p=1000$~MHz.
Arrows indicate the quantum-signal center, the pilot, and the third-order intermodulation replica centered at $2f_p-f_c$.}
\label{fig:replica_position}
\end{figure}

Let $f'$ denote the baseband frequency relative to the quantum-signal center, and let $H(f')$ denote the receiver RRC matched-filter response, whose magnitude equals that of the transmitter pulse-shaping filter.
For an ideal RRC filter, the response is supported on $\mathcal{B}=[-f_B,f_B]$, where $f_B=(1+\beta_{\mathrm{RRC}})R_s/2$ and $\beta_{\mathrm{RRC}}$ is the roll-off factor.
The filter is peak-normalized such that
\begin{equation}
\frac{1}{R_s}
\int_{-F_s/2}^{F_s/2}
|H(f')|^2\,\mathrm{d}f'=1.
\label{eq:h_normalization}
\end{equation}
Because quantization occurs after transmitter pulse shaping, the resulting error is subsequently weighted by the receiver matched-filter response.
The time-averaged quantization-error power at the matched-filter output is therefore
\begin{equation}
\begin{aligned}
P_q
&=
\int_{-f_B}^{f_B}
|H(f')|^2 S_q(f_{\mathrm{c}}+f')
\,\mathrm{d}f'\\
&=
\int_{-f_B}^{f_B}
|H(f')|^2
\left|\mathrm{NTF}_L(f_{\mathrm{c}}+f')\right|^2
S_e(f_{\mathrm{c}}+f')
\,\mathrm{d}f'.
\end{aligned}
\label{eq:inband_noise_general}
\end{equation}

To obtain an analytical SQNR estimate, we adopt a zero-mean white-noise approximation for the internal quantization error \cite{Widrow1996}.
Using $\sigma_e^2$ for the total noise power, the two-sided noise PSD is
\begin{equation}
S_e(f)
\approx
\frac{\sigma_e^2}{F_s},
\qquad
-\frac{F_s}{2}
\leq f<
\frac{F_s}{2}.
\label{eq:white_error_psd}
\end{equation}
Substituting (\ref{eq:white_error_psd}) into (\ref{eq:inband_noise_general}) gives
\begin{equation}
P_q
\approx
\frac{\sigma_e^2}{F_s}
\int_{-f_B}^{f_B}
|H(f')|^2
\left|\mathrm{NTF}_L(f_{\mathrm{c}}+f')\right|^2
\,\mathrm{d}f'.
\label{eq:inband_noise}
\end{equation}

Let $P_s$ denote the quantum signal power at the same matched-filter output.
Assuming each quadrature error is uniformly distributed over an equal interval, $\Delta_I=\Delta_Q=\Delta$, yields
\begin{equation}
\sigma_e^2 =
\frac{\Delta^2}{6}
\ .
\label{eq:complex_qvar}
\end{equation}
Combining (\ref{eq:ntf}), (\ref{eq:inband_noise}) and (\ref{eq:complex_qvar}) yields
\begin{equation}
\mathrm{SQNR}
\equiv
\frac{P_s}{P_q}
\approx
\frac{6 P_s F_s}
{
\Delta^2
\displaystyle\int_{-f_B}^{f_B}
|H(f')|^2
\left[
2\sin\!\left(
\frac{\pi(f_{\mathrm{c}}+f')}{F_s}
\right)
\right]^{2L}
\,\mathrm{d}f'
}.
\label{eq:sqnr_final}
\end{equation}
This expression accounts for the combined effects of DSM order, quantizer resolution, signal center frequency, sampling rate, and receiver filtering.

\begin{table}[!tbp]
\centering
\caption{
Correlation coefficients of out-of-band replicas for Gaussian modulation with uniform quantization and first-order DSM.
}
\label{tab:replica_correlation}
\footnotesize
\setlength{\tabcolsep}{2pt}
\renewcommand{\arraystretch}{1.15}
\begin{tabular*}{\columnwidth}{@{\extracolsep{\fill}}cc*{4}{r}@{}}
\toprule
& & \multicolumn{2}{c}{Uniform}
  & \multicolumn{2}{c}{First-order DSM} \\
\cmidrule(lr){3-4}\cmidrule(lr){5-6}
No. & $f$ (GHz)
  & $\rho_I$ & $\rho_Q$ & $\rho_I$ & $\rho_Q$ \\
\midrule
\textcircled{\scriptsize 1} & $1.1$
  & $-0.84$ & $0.83$ & $1.2\times10^{-4}$ & $-4.2\times10^{-4}$ \\
\textcircled{\scriptsize 2} & $3.1$
  & $0.86$ & $0.86$ & $-3.8\times10^{-4}$ & $-1.9\times10^{-4}$ \\
\textcircled{\scriptsize 3} & $3.9$
  & $-0.66$ & $0.54$ & $-4.3\times10^{-5}$ & $1.3\times10^{-3}$ \\
\textcircled{\scriptsize 4} & $-1.7$
  & $-0.81$ & $0.81$ & $-3.3\times10^{-4}$ & $-8.5\times10^{-5}$ \\
\textcircled{\scriptsize 5} & $-2.5$
  & $0.85$ & $0.84$ & $3.1\times10^{-5}$ & $-3.8\times10^{-4}$ \\
\textcircled{\scriptsize 6} & $-4.5$
  & $-0.58$ & $0.58$ & $3.9\times10^{-4}$ & $7.3\times10^{-4}$ \\
\bottomrule
\end{tabular*}
\par\vspace{3pt}
\begin{minipage}{\columnwidth}
\footnotesize
\textit{Note:} $f$ denotes the replica center frequency.
$\rho_I$ and $\rho_Q$ are the correlation coefficients of the real and imaginary sequences with their corresponding original sequences.
The sampling rate is 12~GHz, and the quantizer resolution is 2 bits.
\end{minipage}
\end{table}

\begin{figure}[!tbp]
\centering
\includegraphics[width=\linewidth]{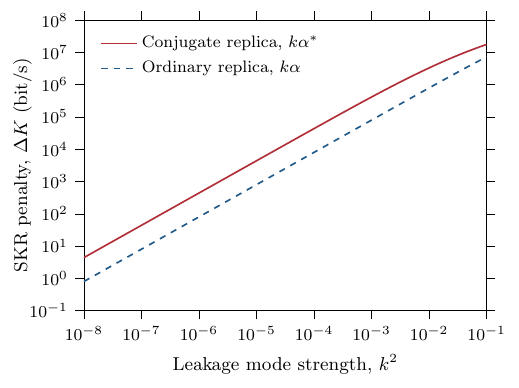}
\caption{
Asymptotic secret key rate penalty.
The conjugate replica considered in our case is compared with the ordinary copy reference~\cite{Jain2021} at the same $k^2$.
The parameters are $V_A=4$, $T=0.355$, $\xi=0.016$, $\eta=0.72$, $v_{\mathrm{el}}=0.076$, $\beta=0.98$, $R_s=500$~MBaud.
}
\label{fig:conjugate_key_penalty}
\end{figure}

\section{Simulation and Spectral Analysis}

Table~\ref{tab:parameters} summarizes the principal simulation and experimental parameters.
The SQNR simulations use $1.024\times10^6$ Gaussian symbols at a symbol rate of 500~MBaud, with a pilot-to-signal power ratio (PSPR) of 10.
To calculate the SQNR, the signal waveforms before and after quantization are downconverted to baseband, then the same matched filter is used to extract the components within the QKD band.

Both OSR and QKD center frequency critically affect the in-band SQNR.
We simulate 3-bit and 8-bit DSMs to verify the impact of high OSR and the QKD center-frequency setting on SQNR, and show the results in Fig.~\ref{fig:sqnr}.
Figure~\ref{fig:sqnr}(a) shows that the in-band SQNR increases with OSR in general, with DSM providing a steeper improvement than uniform quantization.
At sufficiently high OSRs, second-order DSM achieves a significantly higher SQNR than first-order DSM.
At OSR~=~3, however, both DSM schemes perform worse than uniform quantization, and increasing the DSM order further degrades the SQNR.
This is because the DSM, regardless of order, suppresses quantization noise only for $|f|<F_s/6$ within the first Nyquist band.
At OSR~=~3, this boundary is 250~MHz, whereas the quantum-signal band extends from approximately 25 to 575~MHz.
A substantial portion of the signal band therefore lies in the noise-amplification region, and the NTF, weighted by the receiver-filter response, yields higher in-band noise power than uniform quantization, with a greater penalty for second-order DSM.
As OSR increases, the fixed quantum-signal band moves closer to the NTF zero in normalized frequency, enabling increasingly effective in-band noise suppression.
Figure~\ref{fig:sqnr}(b) uses a fixed 12-fold OSR and shows that the DSM is beneficial when the QKD center frequency is smaller than 1~GHz, when a smaller part of the QKD band is inside the DSM noise amplification region.
This result indicates that the noise-suppression boundary at $F_s/6$ should not be treated as a hard limit for QKD band placement; instead, a joint design of the QKD center frequency, sampling rate, and matched-filter response is necessary.
All results agree well with the theoretical curves given by (\ref{eq:sqnr_final}).

We next examine the out-of-band spectral components observed after low-resolution quantization of the composite quantum-signal and pilot waveform.
Figure~\ref{fig:replica}(a) compares a 500-MBaud Gaussian-modulated waveform with a 500-MBaud QPSK test waveform; the latter is included as a diagnostic to clearly demonstrate the data-carrying structure of the spectral replicas.
The simulation uses a sampling rate of 12~GSa/s, a quantum-signal center frequency of 300~MHz, a pilot frequency of 700~MHz, and a PSPR of 10.
For illustration purposes only, Fig.~\ref{fig:replica}(a) compares 2-bit uniform quantization with 2-bit first-order DSM and clearly shows that the quantization produces several distinguishable spectral replicas in addition to the original signal at 300~MHz.

After separate downconversion, matched filtering, and downsampling, the QPSK replica centered at 1.1~GHz shows a recognizable constellation under uniform quantization, shown in the upper panel of Fig.~\ref{fig:replica}(a).
We quantify the Pearson correlation coefficients between the replica and the original signals using Gaussian-modulated symbols and summarize the results in Table~\ref{tab:replica_correlation}.
We find strong same-quadrature linear correlations between corresponding I- and Q-components for uniform quantization, and a first-order DSM significantly reduces these correlations, which is also supported by the noise-like constellation at 1.1~GHz shown in the lower panel of Fig.~\ref{fig:replica}(a).

Figure~\ref {fig:replica}(b) uses Gaussian modulation only and examines the same replica band at 1.1~GHz over the OSR sweep with 3--5-bit quantization.
For uniform quantization, the correlation between signal and replica is pronounced only at 3 bits and is close to zero for 4 bits and above.
With first-order DSM, the estimated correlations remain close to zero across all tested resolutions and OSRs.
DSM therefore changes the structure of the correlated spectral components created by low-resolution quantization and strongly suppresses the linear correlation between the replica and the original symbols.

As shown in Fig.~\ref{fig:replica_position}, increasing the pilot frequency $f_{\mathrm{p}}$ from 0.7 to 1.0~GHz in steps of 0.1~GHz, while keeping $f_{\mathrm{c}}=0.3$~GHz fixed, shifts the replica center frequency from 1.1 to 1.7~GHz in steps of 0.2~GHz.
These frequencies satisfy $f_{\mathrm{replica}}=2f_{\mathrm{p}}-f_{\mathrm{c}}$, and the I- and Q-component correlations have opposite signs, indicating that this replica is a third-order intermodulation product between the pilot tone and the quantum signal. Namely,
\begin{equation}
x_\mathrm{replica}(t) \simeq
C_{\rm IM}
\left(
A_p e^{j2\pi f_p t}
\right)^2
\left(
s(t)e^{j2\pi f_c t}
\right)^*
\ ,
\label{eq:intermodulation_product}
\end{equation}
where $C_{\rm IM}$ is the intermodulation coefficient.

\begin{figure*}[!b]
\centering
\includegraphics[width=0.95\textwidth]{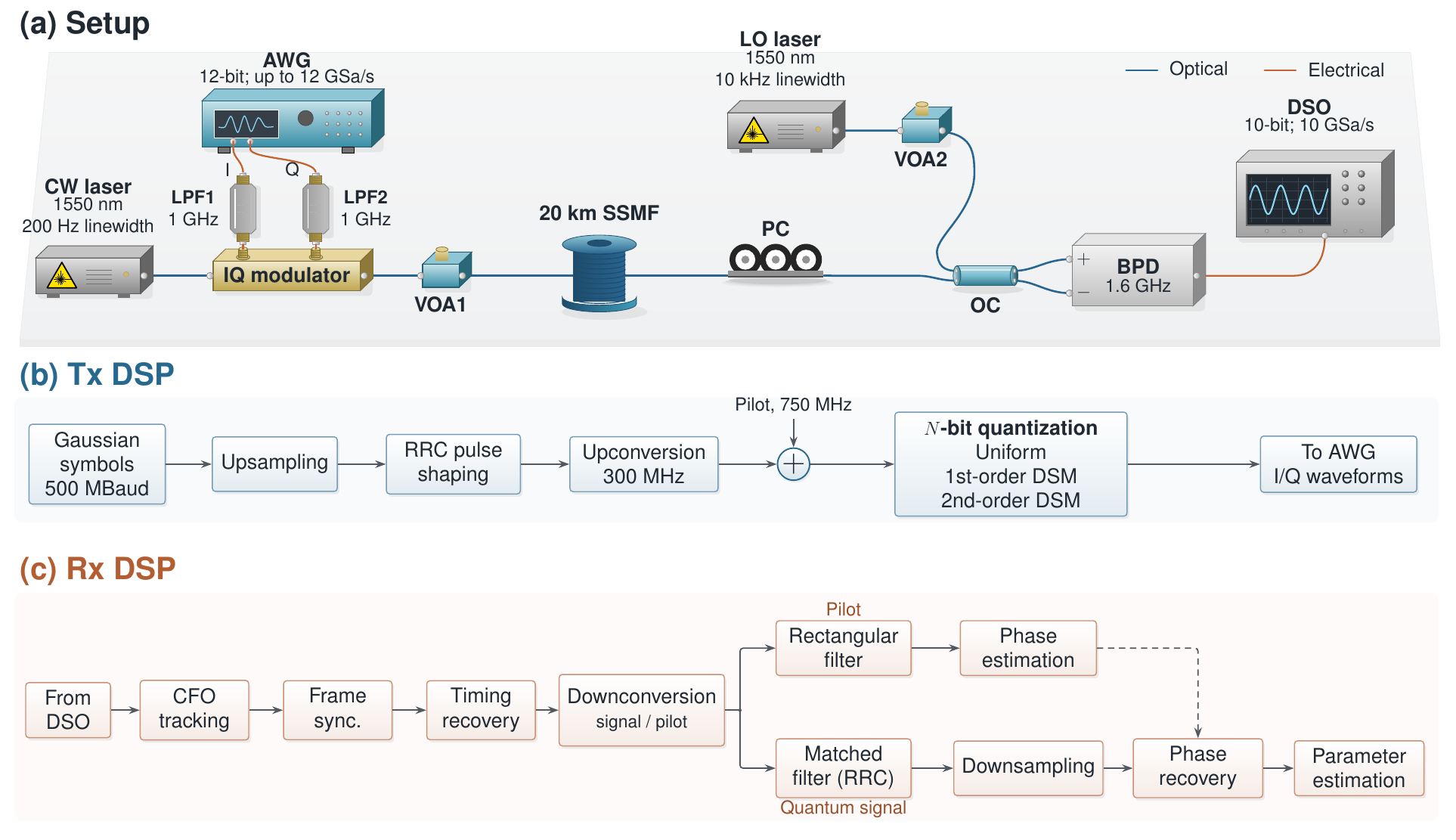}
\caption{
(a) Experimental setup and digital signal processing (DSP) for the experimental Gaussian-modulated CVQKD system.
(b) Transmitter DSP, with one of three quantization schemes selected: $N$-bit uniform quantization, first-order DSM, or second-order DSM. (c) Receiver DSP.
AWG, arbitrary waveform generator; LPF, low-pass filter; IQ Mod, IQ modulator; VOA, variable optical attenuator; PC, polarization controller; OC, optical coupler; BPD, balanced photodetector; DSO, digital storage oscilloscope; CW, continuous wave.
}
\label{fig:setup}
\end{figure*}

\section{Security Analysis}
\label{sec:conjugate_security}

\subsection{Source Model with Conjugate Leakage Mode}

We now relate the measured linear replica component to a source-leakage model, following Jain et al.~\cite{Jain2021}.
We restrict the analysis to the selected replica band at $2f_{\mathrm{p}}-f_{\mathrm{c}}$, since we can always use additional electrical filters to suppress high-frequency replicas.
Let $x_I$ and $x_Q$ be independent Gaussian modulation variables with variance $V_A$ per quadrature, and define the coherent amplitude $\alpha=(x_I+jx_Q)/2$ expressed in shot-noise units (SNU).
We assume a balanced source prepared by Alice with DSM-induced leakage, expressed as
\begin{equation}
|\psi_{\alpha}\rangle_{SL}
=|\alpha\rangle_S\otimes|k\alpha^*\rangle_L,
\qquad k\geq0,
\label{eq:negative_conjugate_source}
\end{equation}
where $S$ is the intended signal, $L$ is the leakage mode, and $k$ is the relative amplitude ratio between replica and signal.
For replica quadratures $y_j$, with $j\in\{I,Q\}$, the $k$ parameter is related to the measured correlation coefficients by
\begin{equation}
\begin{aligned}
k = -\rho_I
\sqrt{\frac{\operatorname{Var}(y_I)}{V_A}}
= \rho_Q
\sqrt{\frac{\operatorname{Var}(y_Q)}{V_A}}
.
\end{aligned}
\label{eq:replica_gain}
\end{equation}
Here $\rho_j$ is the correlation coefficient discussed in Fig.~\ref{fig:replica}.

Using $I_2$ for the identity matrix and $Z=\operatorname{diag}(1,-1)$, the source covariance is
\begin{equation}
{\Gamma}_{SL} =
\begin{pmatrix}
(1+V_A) I_2 & kV_A Z\\
kV_A Z & (1+k^2V_A) I_2
\end{pmatrix}
\ .
\label{eq:replica_source_covariance}
\end{equation}
For comparison, an ordinary leakage mode $|k\alpha\rangle_L$, as used in~\cite{Jain2021}, corresponds to replacing $Z$ with $I_2$ in the covariance matrix.
Both sources, with ordinary or conjugate leakage mode, have the same leakage energy and the same information $\chi(\alpha:L)=g(k^2V_A/2)$, where $g(n)=(n+1)\log_2(n+1)-n\log_2n$.
However, their joint information with a third signal mode need not be equal, because a conjugate state pair can support more informative joint measurements than an ordinary state pair~\cite{Cerf2001PhaseConjugation}.
Consequently, in our case, the joint source with a phase-conjugated leakage mode allows more information leakage to Eve.

\subsection{Holevo Information and SKR Penalties}

We model the attack as an independent thermal-loss channel with transmittance $T$ and input-referred excess noise $\xi$.
Eve holds the channel-environment output $E_1$, the purification $E_2$, and the leakage mode $L$, and has an ideal phase reference for joint processing.
The thermal environment variance is
\begin{equation}
\omega=1+\frac{T\xi}{1-T}.
\label{eq:replica_environment}
\end{equation}
In the mode ordering $E=(E_1,E_2,L)$, the covariance matrix available to Eve is
\begin{equation}
{\Gamma}_{E}
=
\begin{pmatrix}
a I_2 & c Z & d Z\\
c Z & \omega I_2 & 0\\
d Z & 0 & \ell I_2
\end{pmatrix},
\label{eq:replica_eve_covariance}
\end{equation}
where
\begin{equation}
\begin{aligned}
a&=(1-T)(1+V_A)+T\omega,\\
c&=\sqrt{T(\omega^2-1)},\\
d&=-\sqrt{1-T}\,kV_A,\\
\ell&=1+k^2V_A.
\end{aligned}
\label{eq:replica_eve_coefficients}
\end{equation}
Bob's optical covariance is ${\Gamma}_B=b I_2$, with $b=1+T(V_A+\xi)$.
The cross-covariance with Eve is
\begin{equation}
C_{EB}
=
\begin{pmatrix}
\sqrt{T(1-T)}(\omega-1-V_A) I_2\\
\sqrt{(1-T)(\omega^2-1)} Z\\
\sqrt T\,kV_A Z
\end{pmatrix}.
\label{eq:replica_cross_covariance}
\end{equation}

Bob's detector efficiency $\eta$ and electronic noise $v_{\mathrm{el}}$ are trusted receiver parameters.
Using the heterodyne convention in which each detector's normalized electronic-noise contribution is $v_{\mathrm{el}}$, the equivalent measurement noise is
\begin{equation}
\chi_{\mathrm{het}}
=\frac{2-\eta+2v_{\mathrm{el}}}{\eta}.
\label{eq:replica_heterodyne_noise}
\end{equation}
Gaussian conditioning on Bob's measurement gives~\cite{Weedbrook2012}
\begin{equation}
{\Gamma}_{E|B}
={\Gamma}_{E}
-\frac{C_{EB} C_{EB}^{\mathsf T}}
{b+\chi_{\mathrm{het}}}.
\label{eq:replica_conditional_covariance}
\end{equation}
If $\nu_i$ and $\widetilde{\nu}_i$ are the three symplectic eigenvalues of ${\Gamma}_{E}$ and ${\Gamma}_{E|B}$, respectively, the Holevo information with consideration of the DSM-induced leakage mode is
\begin{equation}
\chi_{BE}(k^2)
=\sum_{i=1}^{3}
\left[
g\!\left(\frac{\nu_i-1}{2}\right)
-g\!\left(\frac{\widetilde{\nu}_i-1}{2}\right)
\right].
\label{eq:replica_holevo}
\end{equation}
The Alice--Bob mutual information is
\begin{equation}
I_{AB}
=\log_2\!\left(
\frac{b+\chi_{\mathrm{het}}}
{1+T\xi+\chi_{\mathrm{het}}}
\right).
\label{eq:replica_mutual_information}
\end{equation}
The asymptotic secret key rate (SKR) is given by
\begin{equation}
K(k^2)
= R_s\left[\beta I_{AB}-\chi_{BE}(k^2)\right].
\label{eq:replica_model_rate}
\end{equation}
At fixed $V_A$, $T$, $\xi$, and receiver parameters, Bob's statistics and $I_{AB}$ do not change with $k^2$.
The isolated leakage penalty, in the positive-key regime considered below, is therefore~\cite{Jain2021}
\begin{equation}
\begin{aligned}
\Delta K(k^2)
&=K(0)-K(k^2)
\ .
\end{aligned}
\label{eq:replica_key_penalty}
\end{equation}

Figure~\ref{fig:conjugate_key_penalty} compares the SKR penalties for the ordinary and conjugate replicas.
For $k^2=10^{-8}$, $10^{-7}$, and $10^{-6}$, the SKR penalties due to the conjugate leakage mode are approximately 4.38, 43.8, and 438~bit/s, respectively.
The corresponding ordinary-leakage penalties are 0.797, 7.97, and 79.7~bit/s.
Thus, in this weak-leakage interval, the conjugate-source penalty is approximately 5.5 times the ordinary-copy prediction, although both remain small compared with the final SKR.
The penalty ratio is not universal and depends on channel noise, loss, and leakage strength.

\begin{figure}[!tbp]
\centering
\includegraphics[width=\linewidth]{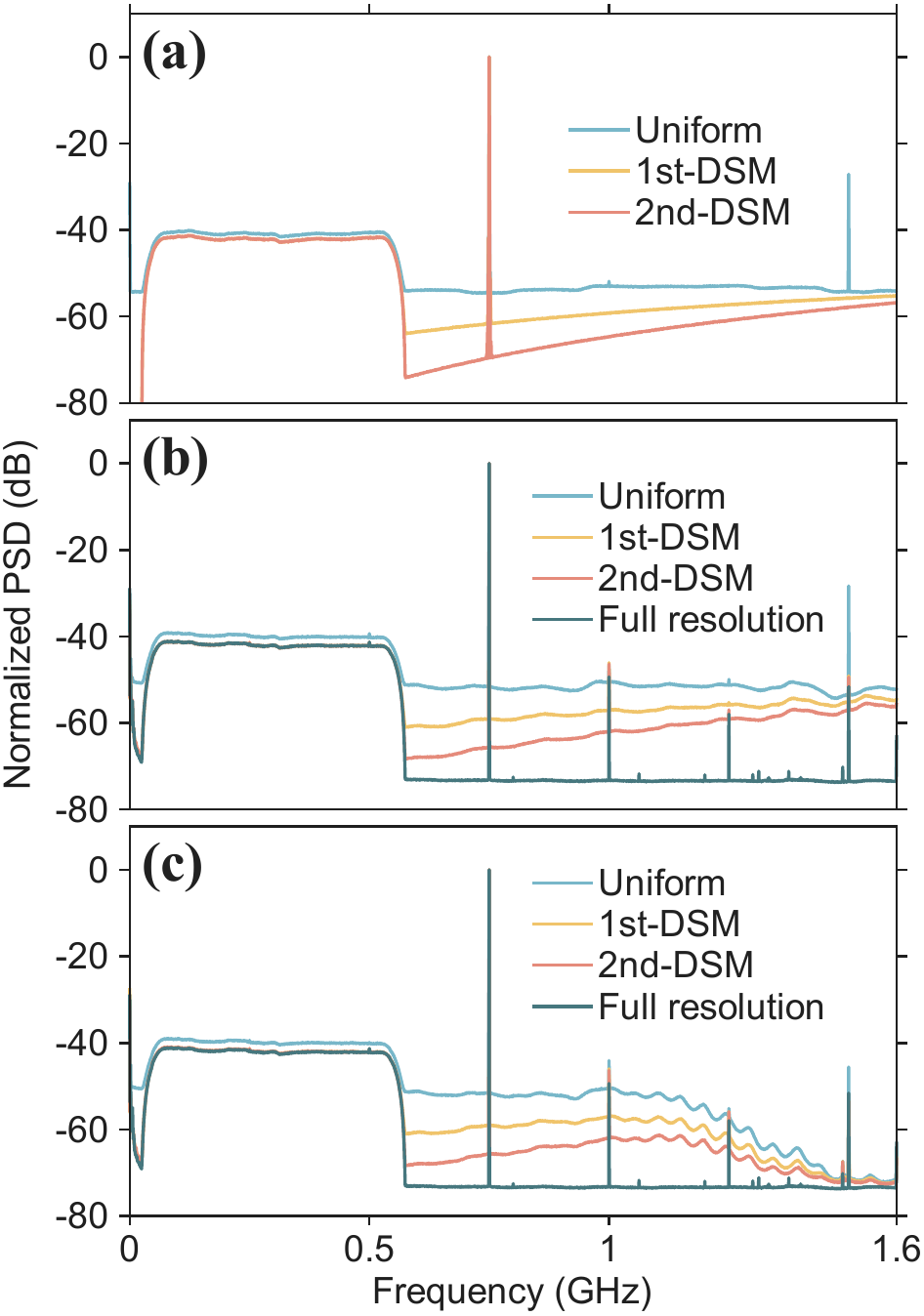}
\caption{
Normalized power spectral densities for 3-bit quantized (a) digital waveform at the AWG input; (b) AWG electrical output captured directly by the oscilloscope; (c) AWG electrical output captured by the oscilloscope after passing through an external 1-GHz low-pass filter.
}
\label{fig:psd}
\end{figure}

\begin{figure}[!tbp]
\centering
\includegraphics[width=\linewidth]{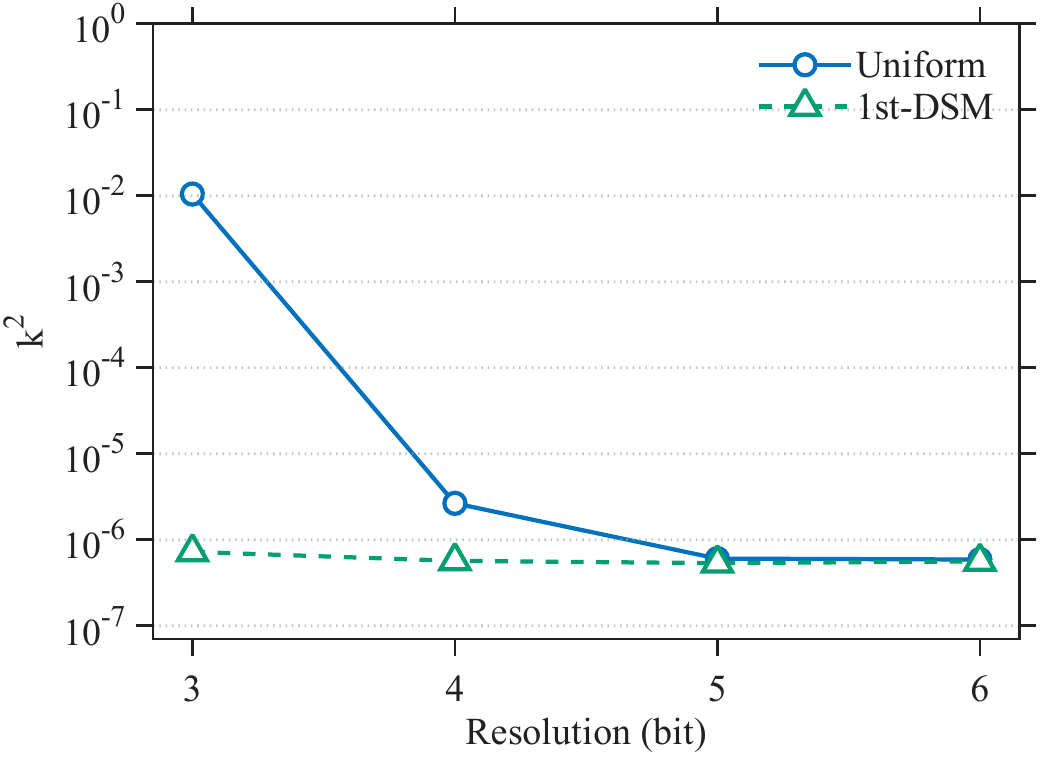}
\caption{
Experimentally estimated leakage-mode strength $k^2$ of the replica centered at 1.2~GHz versus the digitally imposed quantization resolution.
Uniform quantization and first-order DSM are compared at $\mathrm{OSR}=24$ and a PSPR of 10~dB.
The estimates are obtained from optical measurements and referred to Alice's transmitter output.
Each point represents the mean of the I- and Q-quadrature estimates from a single evaluation using $N=10^8$ symbols.
}
\label{fig:experimental_leakage}
\end{figure}

\section{Experimental Setup}

Figure~\ref{fig:setup} shows the experimental Gaussian-modulated CVQKD system with a locally generated local oscillator (LO) at Bob.
Alice uses a continuous-wave (CW) laser with a center wavelength of 1550~nm and a linewidth of 200~Hz as the coherent optical source.
The transmitter first generates independent and identically distributed Gaussian random symbols at a symbol rate of 500~MBaud, then upsamples and pulse-shapes them using an RRC filter with a roll-off factor of 0.1.
The shaped quantum signal is upconverted to 300~MHz and combined with a pilot tone at 750~MHz.
The composite waveform is then digitally quantized using uniform quantization, first-order DSM, or second-order DSM.
An arbitrary waveform generator (AWG) with a nominal 12-bit resolution and a maximum sampling rate of 12~GSa/s generates the electrical signal; the actual sampling rate is adjusted to the selected OSR.
The I and Q outputs pass through external 1-GHz low-pass filters before driving a single-polarization IQ modulator that loads the composite waveform onto the optical carrier.
A variable optical attenuator controls the launch power and sets the quantum-signal modulation variance to $V_A=4$~SNU.
The modulated optical signal is transmitted to Bob through 20~km of standard single-mode fiber.

Bob uses an independent CW laser with a 10~kHz linewidth as the local oscillator.
The received signal and LO are combined in an optical coupler and measured by a 1.6~GHz balanced photodetector (BPD) for heterodyne coherent reception.
A 10-bit oscilloscope sampling at 10~GSa/s captures the BPD electrical output.
The receiver DSP performs carrier frequency offset (CFO) tracking, frame synchronization, and timing recovery, then separately downconverts the quantum and pilot signals to baseband.
The quantum signal is matched-filtered and downsampled, while the pilot signal is filtered, and its phase is extracted for phase recovery~\cite{Liao2025Wiener}.

The asymptotic secret key rate is calculated for reverse reconciliation and heterodyne detection using the Gaussian-modulated coherent-state CVQKD model \cite{DevetakWinter2005, Navascues2006, GarciaPatron2006} with the expression
\begin{equation}
K = \gamma R_s \left(\beta I_{AB} - \chi_{BE}\right),
\label{eq:skr}
\end{equation}
where $\beta$ is the reverse reconciliation efficiency and $\gamma$ is the fraction of Gaussian symbols used for key evaluation.
The experimental parameters are as follows: $\beta=0.98$, $\gamma=0.84$, detection efficiency $\eta=0.72$, electronic noise $v_{\mathrm{el}}=0.07$, and channel transmittance $T=0.355$.
The detection efficiency and electronic noise are treated as trusted receiver parameters, whereas quantization- and DSM-induced noise contributions are included in the untrusted excess noise.
The Holevo information is evaluated with experimentally measured $k^2$ to incorporate the DSM-induced leakage mode as detailed in the last section.

\begin{figure}[!tbp]
\centering
\includegraphics[width=\linewidth]{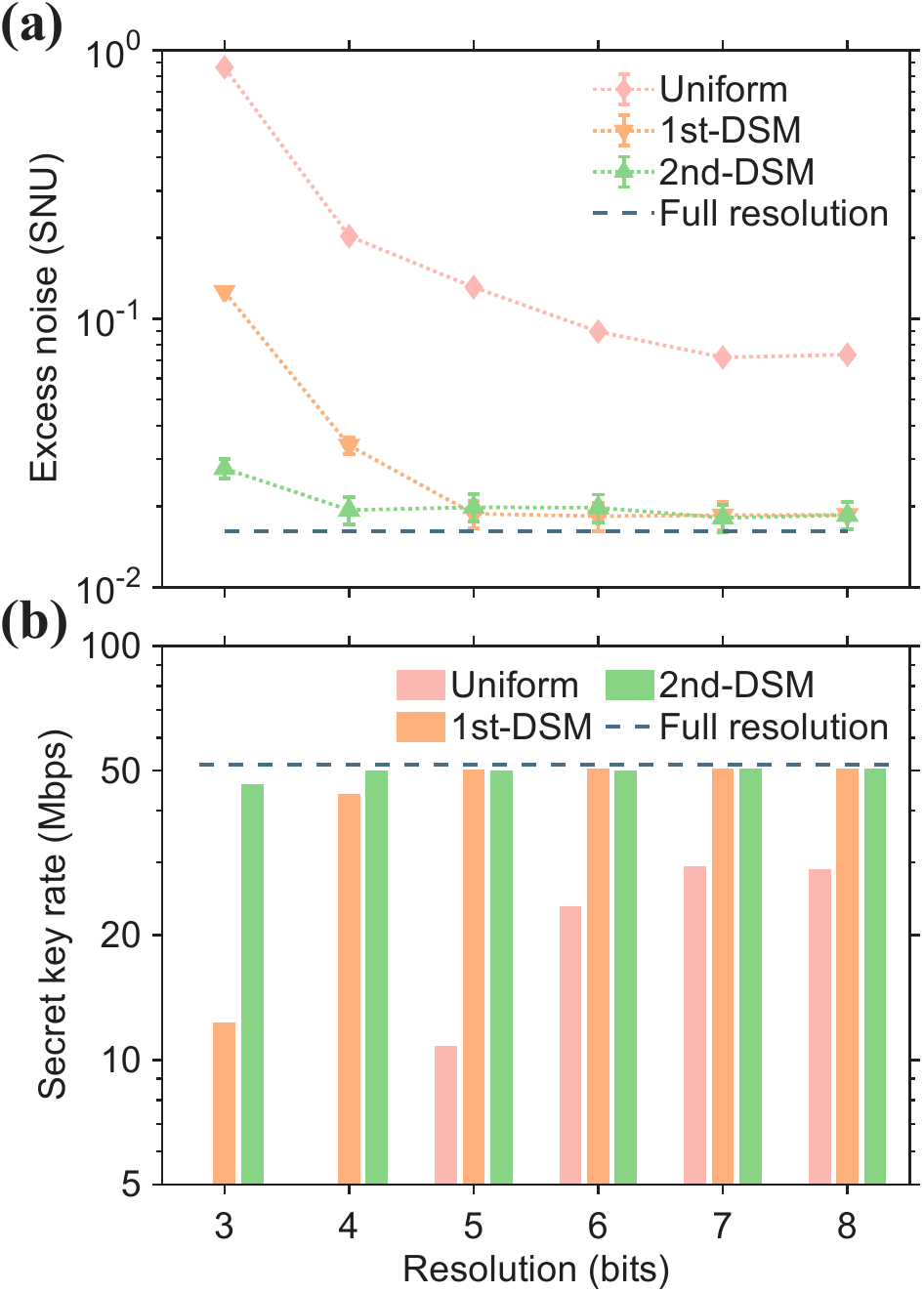}
\caption{
Effect of DSM and quantizer resolution on CVQKD performance, including (a) excess noise referred to Alice’s side; (b) estimated asymptotic key rate at OSR = 12.
The 3- and 4-bit uniform quantization do not yield a positive key rate.
}
\label{fig:resolution}
\end{figure}

\section{Results and Discussion}

\begin{figure*}[!tbp]
\centering
\includegraphics[width=0.95\textwidth]{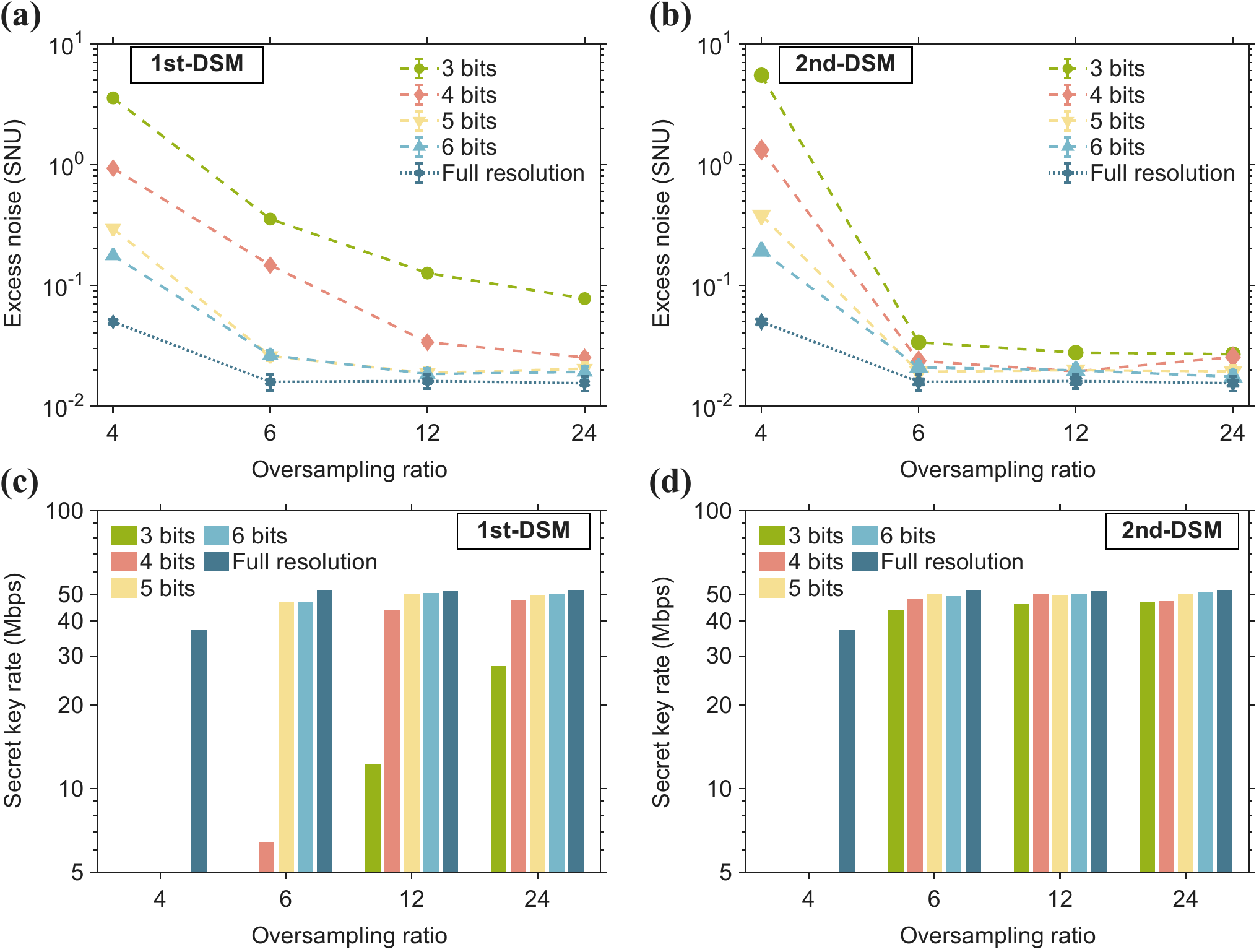}
\caption{
Effects of DSM and quantizer resolution on the GM-LLO-CVQKD performance at different OSRs.
The full resolution is measured separately at each OSR.
Missing key-rate bars indicate that no positive estimate is obtained.
(a) First-order DSM: excess noise referred to Alice’s side; (b) Second-order DSM: excess noise referred to Alice’s side; (c) First-order DSM: estimated key rate; (d) Second-order DSM: estimated key rate.
}
\label{fig:osr_exp}
\end{figure*}

We use a 12~GSa/s AWG to emulate electrical waveforms with various quantization depths and OSRs.
The waveform prepared digitally as double-precision floating-point and quantized directly by the AWG is considered as the full-resolution reference.
To determine whether the DSM noise-shaped spectrum is retained at the physical electrical output, Fig.~\ref{fig:psd} compares the PSD of the AWG-generated 3-bit quantized waveform, with or without DSM, captured directly by the oscilloscope.
Although the oscilloscope bandwidth is 4~GHz, we show only up to 1.6~GHz to match the optical measurements shown below because this is the visible bandwidth of the BPD.
Figure~\ref{fig:psd}(a) shows the PSD of the quantized digital waveform input to the AWG, where the effect of NTF is clearly visible.
The noise-suppression boundary is 2~GHz, 1/6 of the 12~GHz sampling rate.
Figure~\ref{fig:psd}(b) verifies that the AWG generates the designed waveform with high fidelity, the PSDs of which agree well with those shown in (a).
Figure~\ref{fig:psd}(c) shows the PSD of the same waveform at the AWG output after passing through an analog electrical low-pass filter (LPF) with 1~GHz bandwidth.
The LPF strongly suppresses high-frequency quantization noise and DSM-induced replicas.
However, the replica at 1.2~GHz remains, and its variance must be measured as leakage mode to facilitate the SKR calculation.
The characteristic spectral features observed in the digital domain remain identifiable at the electrical output, indicating that the quantization-noise shaping produced by the digital DSM survives the practical AWG output stage.

Figure~\ref{fig:experimental_leakage} experimentally characterizes the leakage-mode strength $k^2$ of the replica centered at 1.2~GHz for uniform quantization and first-order DSM.
The replica quadratures are recovered from optical measurements with the same DSP as the main signal and calibrated to Alice's transmitter output.
The measurements are performed using $10^8$ recovered symbols at each point with $\mathrm{OSR}=24$ and a PSPR of 10~dB.
For 3-bit uniform quantization, the estimated $k^2$ is at the level of $10^{-2}$, indicating an SKR penalty of 3 Mb/s according to Fig.~\ref{fig:conjugate_key_penalty}.
First-order DSM reduces $k^2$ to approximately $7\times10^{-7}$, a four order-of-magnitude reduction.
Increasing the resolution to 4 bits or more substantially reduces replica strength for both uniform quantization and first-order DSM, with negligible SKR penalties.
Thus, under the tested conditions, DSM provides the most pronounced benefit at 3-bit resolution, whereas increasing the quantization resolution substantially reduces the difference between the two methods.

\begin{table}[!htbp]
\centering
\caption{Representative experimental results at OSR~=~12.}
\label{tab:key_results}
\begin{tabular}{@{}lcc@{}}
\toprule
Method & $\xi$ (SNU) & $K$ (Mbps)\\
\midrule
3-bit uniform quantization & 0.864 & No positive value\\
3-bit first-order DSM & 0.126 & 12.28\\
3-bit second-order DSM & 0.028 & 46.38\\
4-bit second-order DSM & 0.019 & 50.11\\
Full resolution & 0.016 & 51.66\\
\bottomrule
\end{tabular}
\end{table}

We continue to study the effects of quantizer resolution and OSR on excess noise and the asymptotic SKR.
Figure~\ref{fig:resolution} presents experimental results at 12-fold OSR, with representative values listed in Table~\ref{tab:key_results}.
Low-resolution quantization without DSM substantially increases the total excess noise and severely reduces the key rate.
For example, 3-bit uniform quantization produces 0.86 SNU excess noise and yields no positive SKR.
First-order DSM lowers the excess noise to 0.13~SNU and restores the key rate to 12.28~Mbps.
Second-order DSM further reduces the excess noise to 0.028~SNU and raises the key rate to 46.38~Mbps.
Hence, a 3-bit DAC running at 12-fold OSR can deliver a positive key rate at a 20~km fiber range, making an appealing case for short-reach applications.
With 4-bit second-order DSM, $\xi$ is 0.019~SNU and the estimated key rate is 50.11~Mbps, close to the full-resolution reference values of 0.016~SNU and 51.66~Mbps.
As the resolution increases, second-order DSM progressively approaches full-resolution performance.

\begin{figure}[!tbp]
\centering
\includegraphics[width=\linewidth]{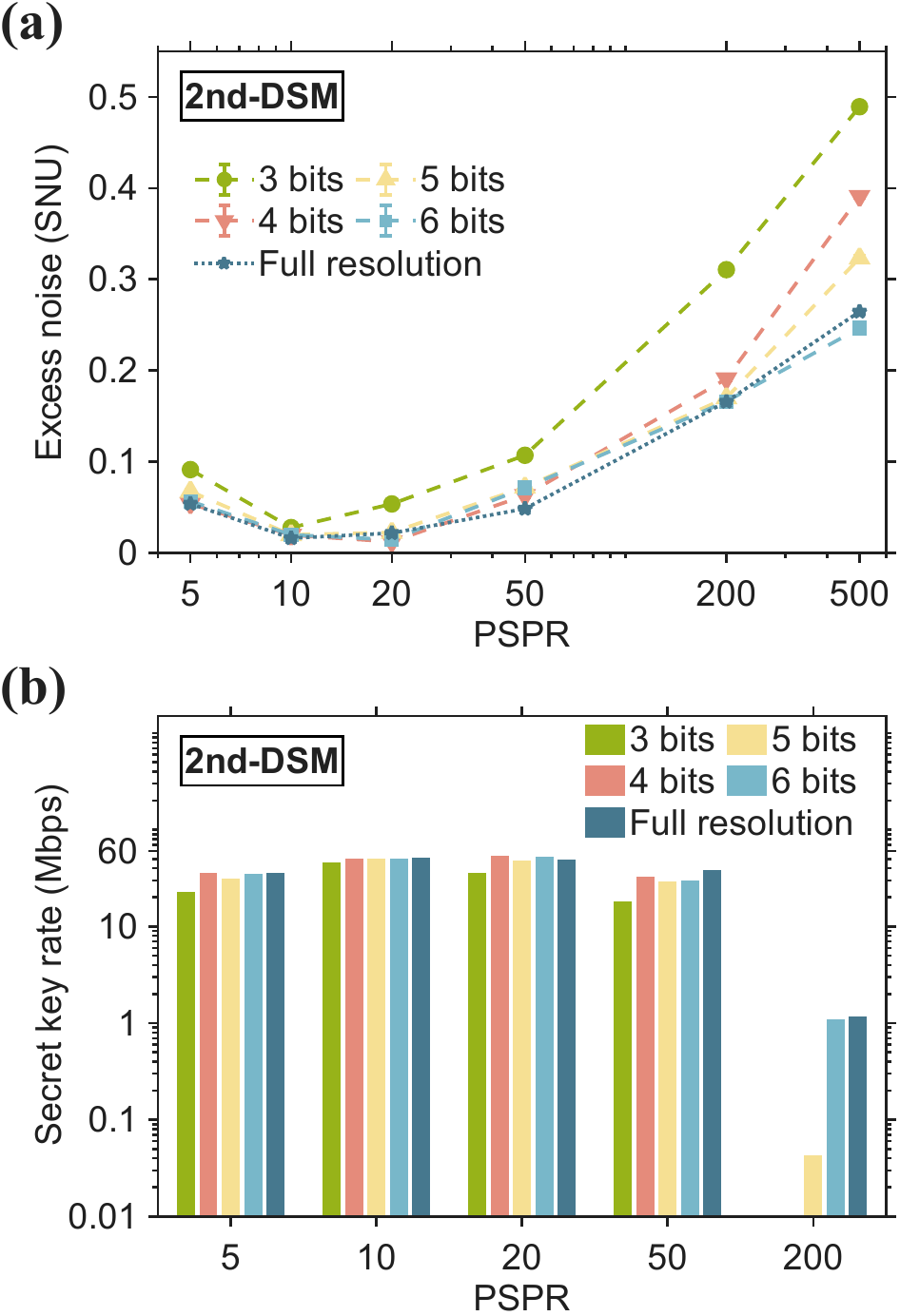}
\caption{
Effect of the linear PSPR and quantizer resolution on the GM-LLO-CVQKD performance of second-order DSM at OSR~=~12.
The quantum-signal power is held fixed while only the power of the 750-MHz pilot is varied.
(a) Excess noise referred to Alice’s side; (b) Estimated asymptotic key rate.
}
\label{fig:pspr}
\end{figure}

Figure~\ref{fig:osr_exp} shows the first- and second-order DSM results at various OSRs.
Both designs working at OSR~=~4 produce too much excess noise and yield no positive key rate with 3--6-bit quantization.
At OSR~=~6 and above, second-order DSM rapidly approaches full-resolution performance at all bit depths considered.
The OSR requirement is more stringent for first-order DSM; even a 24-fold OSR does not recover a substantial key rate with 3-bit quantization.
These results agree well with our analysis, as in Fig.~\ref{fig:sqnr}(a) and ~\eqref{eq:sqnr_final}: at an insufficient sampling rate, the effective QKD band may cross the noise suppression boundary of the NTF, and amplified quantization noise will degrade the overall performance.
Therefore, the quantizer resolution and DSM order must be selected jointly with the OSR, quantum-signal center frequency, and signal bandwidth to optimize system performance.
A second-order DSM generally outperforms a first-order DSM with the parameters considered in this work.
In an engineering design, ~\eqref{eq:sqnr_final} can be evaluated with the implemented RRC response to select the minimum resolution and DSM order consistent with the system noise budget.

Figure~\ref{fig:pspr} shows the experimental results for second-order DSM at OSR~=~12 and various PSPR values.
For the present pilot frequency, quantum-signal power, and receiver DSP, the favorable operating region is approximately PSPR~=~10--20 in linear scale.
Although higher PSPR reduces the quantum-signal dynamic range in the composite waveform and therefore worsens SQNR, laser phase noise dominates the observed performance degradation here.
Random phase noise broadens the pilot spectrum, and its spectral tail causes crosstalk in the QKD band, a typical problem in frequency-multiplexed transmitter designs~\cite {DiDomenico2010}.

\section{Conclusion}

We have investigated transmitter-side delta--sigma noise shaping for low-resolution waveform generation in Gaussian-modulated CVQKD.
By combining the DSM noise-transfer function with the receiver matched-filter response, we developed an in-band quantization-noise model that accounts for quantizer resolution, oversampling ratio, signal center frequency, and DSM order.
For the structures considered here, the DSM suppresses quantization noise within $|f|<F_s/6$, and increasing the DSM order provides stronger suppression.
Therefore, system designers should choose the DAC sampling rate and QKD-band placement jointly so that a substantial part of the quantum-signal band lies within this noise-suppression region.
We identify a nonlinear intermodulation product between the pilot and quantum signal as a phase-conjugated spectral replica that must be considered as a leakage mode during quantum state preparation.
We thus develop a security model that relates this phase-conjugate replica to an additional key-rate penalty.
Simulations and optical measurements show that the DSM strongly suppresses the measured amplitude of quantization-induced replicas, thereby reducing the key-rate penalty.
Experiments conducted on a 500-MBaud CVQKD system over 20~km of optical fiber confirm that, at oversampling rate of 12, a 3-bit second-order DSM restores a positive asymptotic secret key rate of 46.38~Mbps.
With 4-bit second-order DSM, the estimated rate reaches 50.11~Mbps, approximately 97\% of the 51.66~Mbps full-resolution reference.
These results show that suitably designed noise shaping enables low-resolution waveforms to approach full-resolution key-rate performance while mitigating quantization-induced replica leakage within the linear-replica model.
This provides a practical route toward lower-resolution CVQKD transmitters, supported by both an in-band noise model and a quantitative assessment of correlated spectral leakage.

\begin{backmatter}

\bmsection{Funding}
This work was supported in part by the National Natural Science Foundation of China under Grant 62475297 and in part by the Guangdong Basic and Applied Basic Research Foundation under Grant 2024B1515020004.

\bmsection{Disclosures}
The authors declare no conflicts of interest.

\bmsection{Data Availability Statement}
Data underlying the results presented in this paper are not publicly available at this time but may be obtained from the authors upon reasonable request.

\end{backmatter}

\appendix
\section{DSM NTF and Frequency Response}
\label{app:stf_ntf_derivation}

For the unity-coefficient CIFB structure considered here, the $L$th-order state equations under the additive quantization-error model \cite{Gray1987,Pavan2017} are
\begin{equation}
\begin{aligned}
u_m[n]&=u_m[n-1]+u_{m-1}[n]-y[n-1],\\
y[n]&=u_L[n]+e[n],
\end{aligned}
\label{eq:app_dsm_general_states}
\end{equation}
where $u_0[n]=x[n]$, $m=1,\ldots,L$, and $e[n]$ is the internal quantization error.
With zero initial conditions and $d(z)=1-z^{-1}$, the $z$-transformed state equations become
\begin{equation}
d(z)U_m(z)=U_{m-1}(z)-z^{-1}Y(z).
\label{eq:app_dsm_general_state_z}
\end{equation}
Recursively eliminating the internal states and using $U_L(z)=Y(z)-E(z)$ gives
\begin{equation}
d^L(z)\bigl[Y(z)-E(z)\bigr]
=X(z)-z^{-1}Y(z)\sum_{r=0}^{L-1}d^r(z).
\label{eq:app_dsm_general_elimination}
\end{equation}
Since $z^{-1}\sum_{r=0}^{L-1}d^r(z)=1-d^L(z)$, the input-output relation reduces to
\begin{equation}
Y(z)=X(z)+\left(1-z^{-1}\right)^L E(z).
\label{eq:app_dsm_general_output}
\end{equation}
The NTF is therefore
\begin{equation}
\mathrm{NTF}_L(z)
=\left.\frac{Y(z)}{E(z)}\right|_{X(z)=0}
=\left(1-z^{-1}\right)^L.
\label{eq:app_stf_ntf_general}
\end{equation}

Evaluating the NTF on the unit circle $z=e^{j2\pi f/F_{\mathrm{s}}}$ gives its frequency response:
\begin{equation}
\begin{aligned}
\mathrm{NTF}_L(f)
&=\left[2j\,e^{-j\pi f/F_{\mathrm{s}}}
\sin\!\left(\frac{\pi f}{F_{\mathrm{s}}}\right)\right]^L.
\end{aligned}
\label{eq:app_ntf_frequency}
\end{equation}

\FloatBarrier
\bibliography{references}

\begin{thebibliography}{10}
\newcommand{\enquote}[1]{``#1''}

\bibitem{Weedbrook2012}
C.~Weedbrook, S.~Pirandola, R.~Garc{\'i}a-Patr{\'o}n, \emph{et~al.}, \enquote{Gaussian quantum information,} {\protect\JournalTitle{Reviews of Modern Physics}} \textbf{84}, 621--669 (2012).

\bibitem{Laudenbach2018}
F.~Laudenbach \emph{et~al.}, \enquote{Continuous-variable quantum key distribution with gaussian modulation---the theory of practical implementations,} {\protect\JournalTitle{Advanced Quantum Technologies}} \textbf{1}, 1800011 (2018).

\bibitem{Pirandola2020}
S.~Pirandola \emph{et~al.}, \enquote{Advances in quantum cryptography,} {\protect\JournalTitle{Advances in Optics and Photonics}} \textbf{12}, 1012--1236 (2020).

\bibitem{Zhang2019}
G.~Zhang \emph{et~al.}, \enquote{An integrated silicon photonic chip platform for continuous-variable quantum key distribution,} {\protect\JournalTitle{Nature Photonics}} \textbf{13}, 839--842 (2019).

\bibitem{Zhang2024Review}
Y.~Zhang, Y.~Bian, Z.~Li, \emph{et~al.}, \enquote{Continuous-variable quantum key distribution system: Past, present, and future,} {\protect\JournalTitle{Applied Physics Reviews}} \textbf{11}, 011318 (2024).

\bibitem{Wang2024Telecom}
T.~Wang, P.~Huang, L.~Li, \emph{et~al.}, \enquote{High key rate continuous-variable quantum key distribution using telecom optical components,} {\protect\JournalTitle{New Journal of Physics}} \textbf{26}, 023002 (2024).

\bibitem{Hajomer2024}
A.~A.~E. Hajomer, C.~Bruynsteen, I.~Derkach, \emph{et~al.}, \enquote{Continuous-variable quantum key distribution at 10 {GBaud} using an integrated photonic-electronic receiver,} {\protect\JournalTitle{Optica}} \textbf{11}, 1197--1204 (2024).

\bibitem{Hajomer2026Chip16GBaud}
A.~A.~E. Hajomer, A.~Bomhals, C.~Bruynsteen, \emph{et~al.}, \enquote{Chip-based 16 {GBaud} continuous-variable quantum key distribution,} {\protect\JournalTitle{Optica}} \textbf{13}, 1035--1042 (2026).

\bibitem{Liu2025SiliconTRX}
S.~Liu, Y.~Jia, Y.~Shi, \emph{et~al.}, \enquote{Continuous-variable quantum key distribution over 50.4 km fiber using an integrated silicon photonic transmitter and receiver,} {\protect\JournalTitle{Photonics Research}} \textbf{13}, 3141--3150 (2025).

\bibitem{Xu2026GaussianIntegrated}
X.~Xu, L.~Fan, Y.~Pan, \emph{et~al.}, \enquote{Gaussian-modulated continuous-variable quantum key distribution over 60 km fiber using an integrated silicon photonic receiver,} {\protect\JournalTitle{Optics Letters}} \textbf{51}, 3485--3488 (2026).

\bibitem{Pietri2024QOSST}
Y.~Pi{\'e}tri, M.~Schiavon, V.~Marulanda~Acosta, \emph{et~al.}, \enquote{{QOSST}: A highly-modular open source platform for experimental continuous-variable quantum key distribution,} {\protect\JournalTitle{Quantum}} \textbf{8}, 1575 (2024).

\bibitem{Widrow1996}
B.~Widrow, I.~Koll{\'a}r, and M.-C. Liu, \enquote{Statistical theory of quantization,} {\protect\JournalTitle{IEEE Transactions on Instrumentation and Measurement}} \textbf{45}, 353--361 (1996).

\bibitem{Almonacil2020}
S.~Almonacil, F.~Boitier, and P.~Layec, \enquote{Performance model and design rules for optical systems employing low-resolution {DAC}/{ADC},} {\protect\JournalTitle{Journal of Lightwave Technology}} \textbf{38}, 3007--3014 (2020).

\bibitem{Jouguet2012}
P.~Jouguet, S.~Kunz-Jacques, E.~Diamanti, and A.~Leverrier, \enquote{Analysis of imperfections in practical continuous-variable quantum key distribution,} {\protect\JournalTitle{Physical Review A}} \textbf{86}, 032309 (2012).

\bibitem{Kumar2020}
R.~Kumar, X.~Tang, S.~Ren, \emph{et~al.}, \enquote{Performance of continuous variable quantum key distribution system at different detector bandwidth,} {\protect\JournalTitle{Optics Communications}} \textbf{471}, 126034 (2020).

\bibitem{Wang2025Composable}
H.~Wang, Y.~Li, T.~Ye, \emph{et~al.}, \enquote{High-rate continuous-variable quantum key distribution over 100 km fiber with composable security,} {\protect\JournalTitle{Optica}} \textbf{12}, 1657--1667 (2025).

\bibitem{Qi2015}
B.~Qi, P.~Lougovski, R.~Pooser, \emph{et~al.}, \enquote{Generating the local oscillator ``locally'' in continuous-variable quantum key distribution based on coherent detection,} {\protect\JournalTitle{Physical Review X}} \textbf{5}, 041009 (2015).

\bibitem{Kleis2017}
S.~Kleis, M.~Rueckmann, and C.~G. Schaeffer, \enquote{Continuous variable quantum key distribution with a real local oscillator using simultaneous pilot signals,} {\protect\JournalTitle{Optics Letters}} \textbf{42}, 1588--1591 (2017).

\bibitem{Ren2021}
S.~Ren, S.~Yang, A.~Wonfor, \emph{et~al.}, \enquote{Demonstration of high-speed and low-complexity continuous variable quantum key distribution system with local local oscillator,} {\protect\JournalTitle{Scientific Reports}} \textbf{11}, 9454 (2021).

\bibitem{Chin2021CarrierRecovery}
H.-M. Chin, N.~Jain, D.~Zibar, \emph{et~al.}, \enquote{Machine learning aided carrier recovery in continuous-variable quantum key distribution,} {\protect\JournalTitle{npj Quantum Information}} \textbf{7}, 20 (2021).

\bibitem{Sarmiento2025OpticalPilot}
S.~Sarmiento, J.~Tabares, and S.~Etcheverry, \enquote{Low-complexity {CV-QKD} system with optical pilot-tone local oscillator synchronization,} {\protect\JournalTitle{Optics Express}} \textbf{33}, 50707--50717 (2025).

\bibitem{Lin2019}
J.~Lin, T.~Upadhyaya, and N.~L{\"u}tkenhaus, \enquote{Asymptotic security analysis of discrete-modulated continuous-variable quantum key distribution,} {\protect\JournalTitle{Physical Review X}} \textbf{9}, 041064 (2019).

\bibitem{Guo2025BasisEncoding}
M.~Guo, P.~Huang, L.~Huang, \emph{et~al.}, \enquote{Discrete-modulated coherent-state quantum key distribution with basis-encoding,} {\protect\JournalTitle{Research}} \textbf{8}, 0691 (2025).

\bibitem{Pan2026DM126km}
Y.~Pan, M.~Wu, H.~Wang, \emph{et~al.}, \enquote{Long-distance discrete-modulated continuous-variable quantum key distribution over 126.56 km of fiber with a local local oscillator,} {\protect\JournalTitle{Communications Physics}}  (2026). Published online 24 June 2026, \url{https://doi.org/10.1038/s42005-026-02718-0}.

\bibitem{Wang2022}
H.~Wang \emph{et~al.}, \enquote{Sub-{Gbps} key rate four-state continuous-variable quantum key distribution within metropolitan area,} {\protect\JournalTitle{Communications Physics}} \textbf{5}, 162 (2022).

\bibitem{Hajomer2024LLO100km}
A.~A.~E. Hajomer, I.~Derkach, N.~Jain, \emph{et~al.}, \enquote{Long-distance continuous-variable quantum key distribution over 100-km fiber with local local oscillator,} {\protect\JournalTitle{Science Advances}} \textbf{10}, eadi9474 (2024).

\bibitem{Matsenko2026}
S.~Matsenko, A.~Ghazisaeidi, M.~Jarzyna, \emph{et~al.}, \enquote{Machine learning based optimization of {CV-QKD} under practical constraints,} arXiv:2606.31534 [quant-ph] (2026). Preprint.

\bibitem{Gray1993Dither}
R.~M. Gray and T.~G. Stockham, Jr., \enquote{Dithered quantizers,} {\protect\JournalTitle{IEEE Transactions on Information Theory}} \textbf{39}, 805--812 (1993).

\bibitem{Eielsen2017}
A.~A. Eielsen and A.~J. Fleming, \enquote{Improving digital-to-analog converter linearity by large high-frequency dithering,} {\protect\JournalTitle{IEEE Transactions on Circuits and Systems I: Regular Papers}} \textbf{64}, 1409--1420 (2017).

\bibitem{Derkach2017}
I.~Derkach, V.~C. Usenko, and R.~Filip, \enquote{Continuous-variable quantum key distribution with a leakage from state preparation,} {\protect\JournalTitle{Physical Review A}} \textbf{96}, 062309 (2017).

\bibitem{Jain2021}
N.~Jain, I.~Derkach, H.-M. Chin, \emph{et~al.}, \enquote{Modulation leakage vulnerability in continuous-variable quantum key distribution,} {\protect\JournalTitle{Quantum Science and Technology}} \textbf{6}, 045001 (2021).

\bibitem{Hajomer2022}
A.~A.~E. Hajomer, N.~Jain, H.~Mani, \emph{et~al.}, \enquote{Modulation leakage-free continuous-variable quantum key distribution,} {\protect\JournalTitle{npj Quantum Information}} \textbf{8}, 136 (2022).

\bibitem{Liu2026}
X.~Liu, J.~Sun, T.~Wang, \emph{et~al.}, \enquote{Continuous-variable quantum key distribution with a leakage from digital-to-analog conversion,} {\protect\JournalTitle{npj Quantum Information}}  (2026). Published online 25 June 2026, \url{https://doi.org/10.1038/s41534-026-01309-9}.

\bibitem{Inose1962}
H.~Inose, Y.~Yasuda, and J.~Murakami, \enquote{A telemetering system by code modulation---$\delta$-$\sigma$ modulation,} {\protect\JournalTitle{IRE Transactions on Space Electronics and Telemetry}} \textbf{SET-8}, 204--209 (1962).

\bibitem{Gray1987}
R.~M. Gray, \enquote{Oversampled sigma-delta modulation,} {\protect\JournalTitle{IEEE Transactions on Communications}} \textbf{35}, 481--489 (1987).

\bibitem{Pavan2017}
S.~Pavan, R.~Schreier, and G.~C. Temes, \emph{Understanding Delta-Sigma Data Converters} (Wiley-IEEE Press, 2017), 2nd ed.

\bibitem{Wang2019DSM}
J.~Wang, Z.~Jia, L.~A. Campos, and C.~Knittle, \enquote{Delta-sigma modulation for next generation fronthaul interface,} {\protect\JournalTitle{Journal of Lightwave Technology}} \textbf{37}, 2838--2850 (2019).

\bibitem{Zeng2023}
X.~Zeng, M.~Fu, H.~Jiang, \emph{et~al.}, \enquote{Digital pre-distortion based on delta sigma modulation assisted look-up table for optical transmission,} {\protect\JournalTitle{Journal of Lightwave Technology}} \textbf{41}, 5827--5833 (2023).

\bibitem{Xie2024}
T.~Xie, X.~Xin, L.~Fang, \emph{et~al.}, \enquote{{SC} and {OFDM} hybrid coherent optical transmission scheme based on 1-bit bandpass delta-sigma modulation,} {\protect\JournalTitle{Optics Express}} \textbf{32}, 11337--11345 (2024).

\bibitem{Yang2025}
X.~Yang, J.~Yu, L.~Zhang, \emph{et~al.}, \enquote{High spectral efficiency 65536 {QAM} mobile fronthaul based on filtered delta-sigma modulation scheme,} {\protect\JournalTitle{Journal of Lightwave Technology}} \textbf{43}, 4546--4554 (2025).

\bibitem{Zhong2025}
L.~Zhong, Y.~Zhao, S.~Zhang, \emph{et~al.}, \enquote{Channel-response-aware delta-sigma modulator design for low-resolution {DAC}-based pre-equalization systems,} {\protect\JournalTitle{Optics Letters}} \textbf{50}, 6875--6878 (2025).

\bibitem{Chen2023ContinuousMode}
Z.~Chen, X.~Wang, S.~Yu, \emph{et~al.}, \enquote{Continuous-mode quantum key distribution with digital signal processing,} {\protect\JournalTitle{npj Quantum Information}} \textbf{9}, 28 (2023).

\bibitem{Fan2026OECC}
W.~Fan, Z.~Zhao, X.~Yi, and D.~Wang, \enquote{Quantization noise shaping in {CVQKD} using delta-sigma modulation,} in \emph{2026 31st OptoElectronics and Communications Conference (OECC),}  (2026). Paper P3-65.

\bibitem{Pi2023LLO100km}
Y.~Pi, H.~Wang, Y.~Pan, \emph{et~al.}, \enquote{Sub-mbps key-rate continuous-variable quantum key distribution with local local oscillator over 100-km fiber,} {\protect\JournalTitle{Optics Letters}} \textbf{48}, 1766--1769 (2023).

\bibitem{Cerf2001PhaseConjugation}
N.~J. Cerf and S.~Iblisdir, \enquote{Phase conjugation of continuous quantum variables,} {\protect\JournalTitle{Physical Review A}} \textbf{64}, 032307 (2001).

\bibitem{Liao2025Wiener}
J.~Liao, W.~Fan, J.~Huang, \emph{et~al.}, \enquote{Simple phase recovery using multi-pilot tones and wiener filters for continuous variable quantum key distribution,} {\protect\JournalTitle{Optics Express}} \textbf{33}, 35999--36014 (2025).

\bibitem{DevetakWinter2005}
I.~Devetak and A.~Winter, \enquote{Distillation of secret key and entanglement from quantum states,} {\protect\JournalTitle{Proceedings of the Royal Society A: Mathematical, Physical and Engineering Sciences}} \textbf{461}, 207--235 (2005).

\bibitem{Navascues2006}
M.~Navascu{\'e}s, F.~Grosshans, and A.~Ac{\'i}n, \enquote{Optimality of gaussian attacks in continuous-variable quantum cryptography,} {\protect\JournalTitle{Physical Review Letters}} \textbf{97}, 190502 (2006).

\bibitem{GarciaPatron2006}
R.~Garc{\'i}a-Patr{\'o}n and N.~J. Cerf, \enquote{Unconditional optimality of gaussian attacks against continuous-variable quantum key distribution,} {\protect\JournalTitle{Physical Review Letters}} \textbf{97}, 190503 (2006).

\bibitem{DiDomenico2010}
G.~Di~Domenico, S.~Schilt, and P.~Thomann, \enquote{Simple approach to the relation between laser frequency noise and laser line shape,} {\protect\JournalTitle{Applied Optics}} \textbf{49}, 4801--4807 (2010).

\end{thebibliography}

\end{document}